\documentclass{article}

\usepackage{arxiv}

\usepackage[utf8]{inputenc} % allow utf-8 input
\usepackage[T1]{fontenc}    % use 8-bit T1 fonts
\usepackage{hyperref}       % hyperlinks
\usepackage{url}            % simple URL typesetting
\usepackage{booktabs}       % professional-quality tables
\usepackage{amsfonts}       % blackboard math symbols
\usepackage{nicefrac}       % compact symbols for 1/2, etc.
\usepackage{microtype}      % microtypography
\usepackage{graphicx}
\usepackage{doi}

\usepackage{float}

\newcommand{\bit}{\begin{itemize}}
\newcommand{\eit}{\end{itemize}}
\newcommand{\be}{\begin{equation}}
\newcommand{\ee}  {\end{equation}}
\newcommand{\bd}{\begin{displaymath}}
\newcommand{\ed}{  \end{displaymath}}
\newcommand{\bc}{\begin{center}}
\newcommand{\ec}{\end{center}}

\title{On initiation of detonation in large fuel-air clouds}

\author{ \href{https://orcid.org/0000-0002-3155-7093}{\includegraphics[scale=0.06]{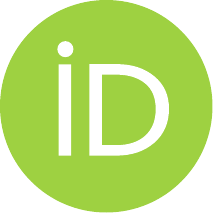}\hspace{1mm}Leonid Kagan}\thanks{Corresponding Author.} \\
Sackler Faculty of Exact Sciences,\\
  School of Mathematical Sciences,\\
  Tel Aviv University, Tel Aviv 69978, Israel\\
  \texttt{kaganleo@tauex.tau.ac.il} \\
  \And
  \href{https://orcid.org/0000-0000-0000-0000}{\includegraphics[scale=0.06]{orcid.pdf}\hspace{1mm}Peter V. Gordon} \\
    Department of Mathematical Sciences,\\
    Kent State University, Ohio 44343, USA\\
    \texttt{peter.gordon@kent.edu} \\
    \And
\href{https://orcid.org/0000-0000-0000-0000}{\includegraphics[scale=0.06]{orcid.pdf}\hspace{1mm}Gregory Sivashinsky} \\
  Sackler Faculty of Exact Sciences,\\
  School of Mathematical Sciences,\\
  Tel Aviv University, Tel Aviv 69978,  Israel\\
  \texttt{grishas@tauex.tau.ac.il} \\       
}

\date{}

\renewcommand{\shorttitle}{On initiation of detonation in large fuel-air clouds}%

\hypersetup{
pdftitle={On initiation of detonation in large fuel-air clouds},
pdfsubject={ArXiv},
pdfauthor={L. Kagan, P.V. Gordon, G. Sivashinsky},
pdfkeywords={Deflagration; Detonation; Deflagration-to-detonation transition;
  Fuel-air clouds},
}

\begin{document}
\maketitle

\begin{abstract}
  The proposed study is motivated by experimental evidence, dating back to 1985,
  demonstrating the
  possibility of deflagration-to-detonation transition (DDT) in a fuel-air
  cloud.  The detonation is initiated by a flame jet developed in a thin
  open-ended tube inserted into the cloud.  Despite the experimental data, a
  first-principle understanding of the mechanism controlling the transition
  is still missing.  
  The current research is aimed at resolution of this issue through a simple
  2D formulation involving minimum physical ingredients.
\end{abstract}

% keywords can be removed

%% main text
\section{Introduction}
\label{sec1}
As is well-known large energies are required for direct initiation of
detonation by an energy source in unconfined fuel-air clouds [1].
Consequently, this scenario is typically excluded in hazard assessments.
However, as has been shown experimentally in 1985 by Moen and his
coworkers [2,3], one can easily initiate detonation employing an
open-ended tube inserted in the cloud.
Nevertheless, attempts to reproduce this mechanism of the DDT initiation
numerically have been unsuccessul up to now [3,4].
The negative outcome of these attempts is a direct consequence of the
$k-\varepsilon$ model employed involving chemistry-independent
turbulence-controlled reaction rate [5,6].
The model predicts the turbulent flame acceleration but not the transition.
However, theoretical results discussed in [7] show that in hydraulically
resisted systems
the turbulence does not play a crucial role in the initiation of DDT,
though chemical kinetics is very important.

The goal of this paper is to show that one can successfully reproduce the
Moen effect within a simple turbulence-free formulation.

Theoretical problem addressed in this work was mentioned by
Li\~{n}an \& Williams in their 1993 book [8] as a problem of detonation
theory meriting attention.

\section{Formulation}

To visulaize the spatial picture of the transition a set of conventional 2D
Navier-Stokes equations for a compressible reactive flow is employed [9].
The reaction rate is modeled by a single-step Arrhenius kinetics.
The latter is assumed to be of the first-order with respect to the deficient
reactant and of the second-order with respect to density (bimolecular reaction).
In suitably chosen units, the set of governing equations reads, 

{\it continuity and state},
\be                                   \label{cs}
\frac{\partial{\hat{\rho}}}{\partial{\hat{t}}}+
\frac{\partial{\hat{\rho}\hat{u}}}{\partial{\hat{x}}}+
\frac{\partial{\hat{\rho}\hat{v}}}{\partial{\hat{y}}}=0,
\hspace*{1.5cm} \hat{P}=\hat{\rho}\hat{T},
%\hspace*{4.5cm}
\ee

{\it momentum},
\be					\label{mx}
\hat{\rho}\left(\frac{\partial{\hat{u}}}{\partial{\hat{t}}}+
\hat{u}\frac{\partial{\hat{u}}}{\partial{\hat{x}}}+
\hat{v}\frac{\partial{\hat{u}}}{\partial{\hat{y}}}\right)+
\frac{1}{\gamma}\frac{\partial{\hat{P}}}{\partial{\hat{x}}}=
%\hspace*{6cm}
%\ee
%\bd
%\hspace*{4cm}
\epsilon Pr \left[
2\frac{\partial{}^2\hat{u}}{\partial{\hat{x}}^2}+
\frac{\partial{}}{\partial{\hat{y}}}\left(
\frac{\partial{\hat{u}}}{\partial{\hat{y}}}+
\frac{\partial{\hat{v}}}{\partial{\hat{x}}}\right)
-\frac{2}{3}\frac{\partial{}}{\partial{\hat{x}}}
\left(\frac{\partial{\hat{u}}}{\partial{\hat{x}}}+
\frac{\partial{\hat{v}}}{\partial{\hat{y}}}\right)
\right],
%\ed
\ee

\be					\label{my}
\hat{\rho}\left(\frac{\partial{\hat{v}}}{\partial{\hat{t}}}+
\hat{u}\frac{\partial{\hat{v}}}{\partial{\hat{x}}}+
\hat{v}\frac{\partial{\hat{v}}}{\partial{\hat{y}}}\right)+
\frac{1}{\gamma}\frac{\partial{\hat{P}}}{\partial{\hat{y}}}=
%\hspace*{6cm}
%\ee
%\bd
%\hspace*{4cm}
\epsilon Pr \left[
2\frac{\partial{}^2\hat{v}}{\partial{\hat{y}}^2}+
\frac{\partial{}}{\partial{\hat{x}}}\left(
\frac{\partial{\hat{u}}}{\partial{\hat{y}}}+
\frac{\partial{\hat{v}}}{\partial{\hat{x}}}\right)
-\frac{2}{3}\frac{\partial{}}{\partial{\hat{y}}}
\left(\frac{\partial{\hat{u}}}{\partial{\hat{x}}}+
\frac{\partial{\hat{v}}}{\partial{\hat{y}}}\right)
\right],
%\ed
\ee

{\it heat},
\be
%\bd							\label{h}
\hat{\rho}\left(\frac{\partial{\hat{T}}}{\partial{\hat{t}}}+
\hat{u}\frac{\partial{\hat{T}}}{\partial{\hat{x}}}+
\hat{v}\frac{\partial{\hat{T}}}{\partial{\hat{y}}}\right)+
%\ed
%\be
(\gamma-1)\hat{P}
\left(\frac{\partial{\hat{u}}}{\partial{\hat{x}}}+
\frac{\partial{\hat{v}}}{\partial{\hat{y}}}\right)=
%\ee
%\bd
\gamma \epsilon \left(\frac{\partial{}^2\hat{T}}{\partial{\hat{x}^2}}+
\frac{\partial{}^2\hat{T}}{\partial{\hat{y}^2}}\right)
+\gamma (\gamma-1) \epsilon Pr \hat{\Phi} + \gamma (1-\sigma_p)\hat{W}, 
%\ed
\ee

where
\be
\hat{\Phi}=2\left(\frac{\partial{\hat{u}}}{\partial{\hat{x}}}\right)^2+
2\left(\frac{\partial{\hat{v}}}{\partial{\hat{y}}}\right)^2+
\left(\frac{\partial{\hat{v}}}{\partial{\hat{x}}}+
\frac{\partial{\hat{u}}}{\partial{\hat{y}}}\right)^2-
%\ee
%\bd
\frac{2}{3}\left(\frac{\partial{\hat{u}}}{\partial{\hat{x}}}+
\frac{\partial{\hat{v}}}{\partial{\hat{y}}}\right)^2,
%\ed
\ee

{\it concentration},                               \label{c}
\be
\hat{\rho}\left(\frac{\partial{\hat{C}}}{\partial{\hat{t}}}+
\hat{u}\frac{\partial{\hat{C}}}{\partial{\hat{x}}}+
\hat{v}\frac{\partial{\hat{C}}}{\partial{\hat{y}}}\right)=
%\ee
%\bd
\frac{\epsilon}{Le}\left(
\frac{\partial{}^2\hat{C}}{\partial{\hat{x}^2}}+
\frac{\partial{}^2\hat{C}}{\partial{\hat{y}^2}}
\right)-\hat{W},%\hspace*{3cm}
%\ed
\ee

{\it chemical kinetics},

\be                                \label{ck}
\hat{W}=Z \hat{\rho}^2\hat{C} \exp \left(N_p(1-\hat{T}^{-1})\right)
\ee

Here $\hat{P}=P/P_0$ is the scaled pressure in units of the initial 
pressure, $P_0$;
$\hat{C}=C/C_0$, scaled concentration of the deficient reactant in units of its initial value, $C_0$;
$\hat{T}=T/T_p$, scaled temperature in units of $T_p=T_0+QC_0/c_p$, adiabatic temperature of burned gas under constant pressure, $P_0$;
$T_0$ is the initial temperature of unburned gas;
$Q$, heat release;
$\sigma_p=T_0/T_p$;
$\gamma=c_p/c_v$; $c_p$, $c_v$, specific heats;
$\epsilon=(u_p/a_p)^2$, scaled thermal diffusivity, where $u_p$, velocity of
the free-space (isobaric) deflagration relative to the burned gas 
is regarded as a prescribed parameter;
$a_p=\sqrt{\gamma(c_p-c_v)T_p}$, sonic velocity at $T=T_p$;
$(\hat{u},\hat{v})=(u,v)/a_p$, scaled flow velocity; 
$N_p=E/RT_p$, scaled activation energy;
$\hat{\rho}=\rho/\rho_p$, where
$\rho_p=P_0/(c_p-c_v)T_p$,
density of the combustion products in free-space deflagration;
$\hat{t}=t/t_p$, 
$(\hat{x},\hat{y})=(x,y)/x_p$, 
$x_p=a_pt_p$, where 
$t_p=A^{-1}Z\exp{(N_p)}$ is the reference time; 
$Z=\frac{1}{2} Le^{-1}N_p^2(1-\sigma_p)^2$ is the
normalizing factor to ensure that at $N_p>>1$ and adiabatic free-space 
conditions
the scaled deflagration velocity relative to the burned gas 
is close to $\sqrt{\epsilon}$;
$A$, pre-exponential factor;
$Pr$, $Le$ - are the Prandtl and Lewis numbers, respectively.
In the adopted formulation the molecular transport coefficients as well as
specific heats are assumed to be constant.

\begin{figure}[!h]
\centering
\includegraphics[scale=0.55]{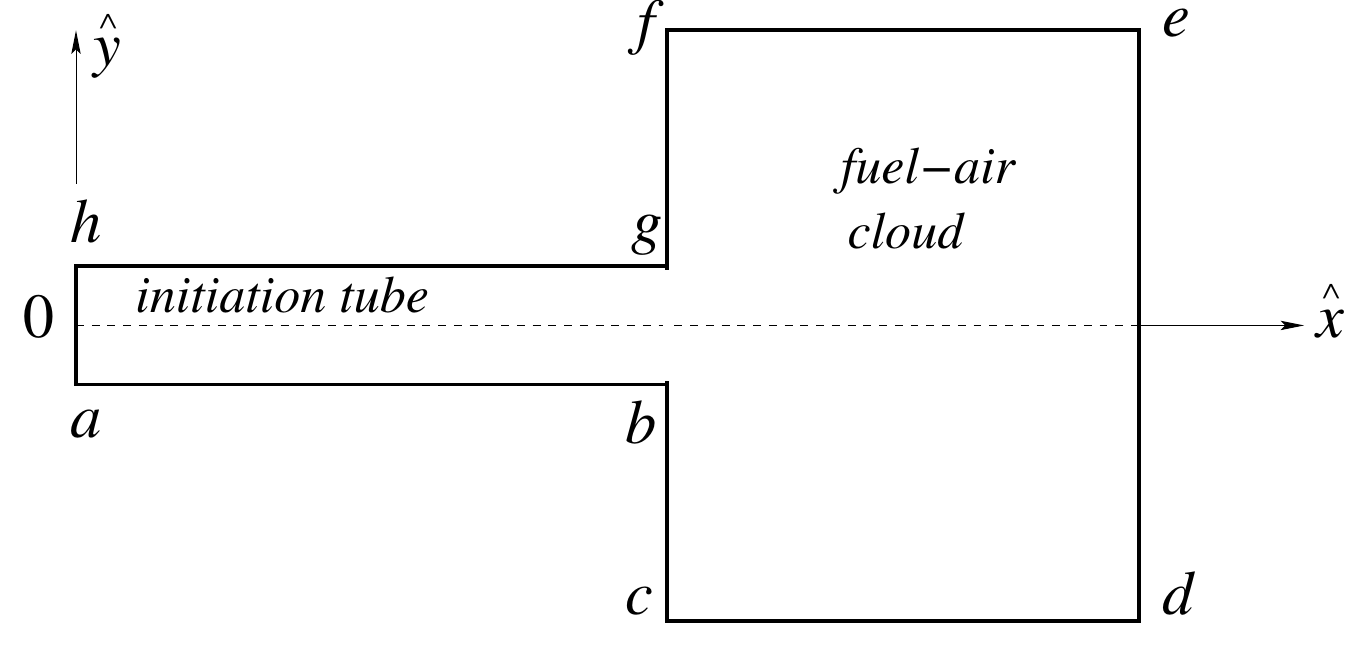}
\caption{
Sketch illustrating the computational area employed. Dashed line marks the system centreline, along which $\partial{\hat{u}}/\partial{\hat{y}}=0$, $\hat{v}=0$, $\partial{\hat{T}}/\partial{\hat{y}}=0$, $\partial{\hat{C}}/\partial{\hat{y}}=0$.}
\end{figure}

As may be readily shown, $x_p=a_pt_p=l_{th}/\sqrt{\epsilon}$, where
$l_{th}=D^0_{th}/u_0$ is the flame width,
$D^0_{th}$, thermal diffusivity at $T=T_0$ and
$u_0=\sigma_p u_p$, velocity of the free-space deflagration relative to 
the fresh mixture.\\

Equations (1)-(7) are considered over the area depicted in (Fig.1) and
suggested by the experimental setup [2] [3].
Here non-slip conditions for the flow velocity ($\hat{u},\hat{v}$) and zero
normal
gradient conditions for the temperature ($\hat{T}$) and concentration
($\hat{C}$)
are held along the boundary  $ah$, $hg$, $gf$, $fe$, $ed$, $dc$, $ch$, $ba$.
The intervals $ah$, $hg$, lengths [$ah$] [$hg$] are fixed while
[$cf$] = 2[$fe$] = 2 [$cd$]  are permanently expanded as the disturbances
(shock and deonation waves) come close to the boundaries $fe$, $ed$, $cd$.

The initial conditions are specified as 
\be
%\bd                                                    \label{ic}
\hat{T}(\hat{x},\hat{y},0)=\sigma_p+(1-\sigma_p)\exp(-\hat{x}/\hat{l}),\;\;
%\ed
%\be
\hat{C}(\hat{x},\hat{y},0)=1,\;\;
\hat{P}(\hat{x},\hat{y},0)=1,\;\;
%\ee
%\bd
\hat{u}(\hat{x},\hat{y},0)=0,\;\;
\hat{v}(\hat{x},\hat{y},0)=0.
%\ed
\ee

Parameter are specified as, $N_p=3.5,4$; $\gamma=1.3,1.4$; $\hat{l}=0.1$;
$\sigma_p=0.2$; $Le=1$; $Pr=0.75$; $\varepsilon=0.01$; $[ah]=15$; $[ab]=36$. 

\section{Numerical simulations}

\begin{figure}[!h]
\centering
\includegraphics[scale=0.55]{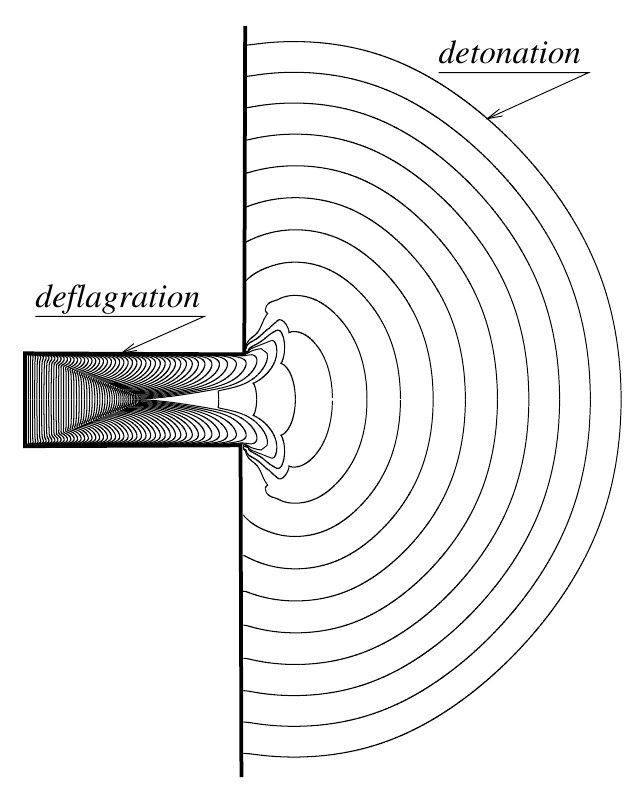}
\caption{
Reaction front configurations illustrating deflagration-to-detonation
transition triggered by a flame jet developing in a thin tube inserted into a
semi-infinite cloud,  $N_p = 3.5$, $\gamma= 1.3$.}
\end{figure}

\begin{figure}[!h]
\centering
\includegraphics[scale=0.55]{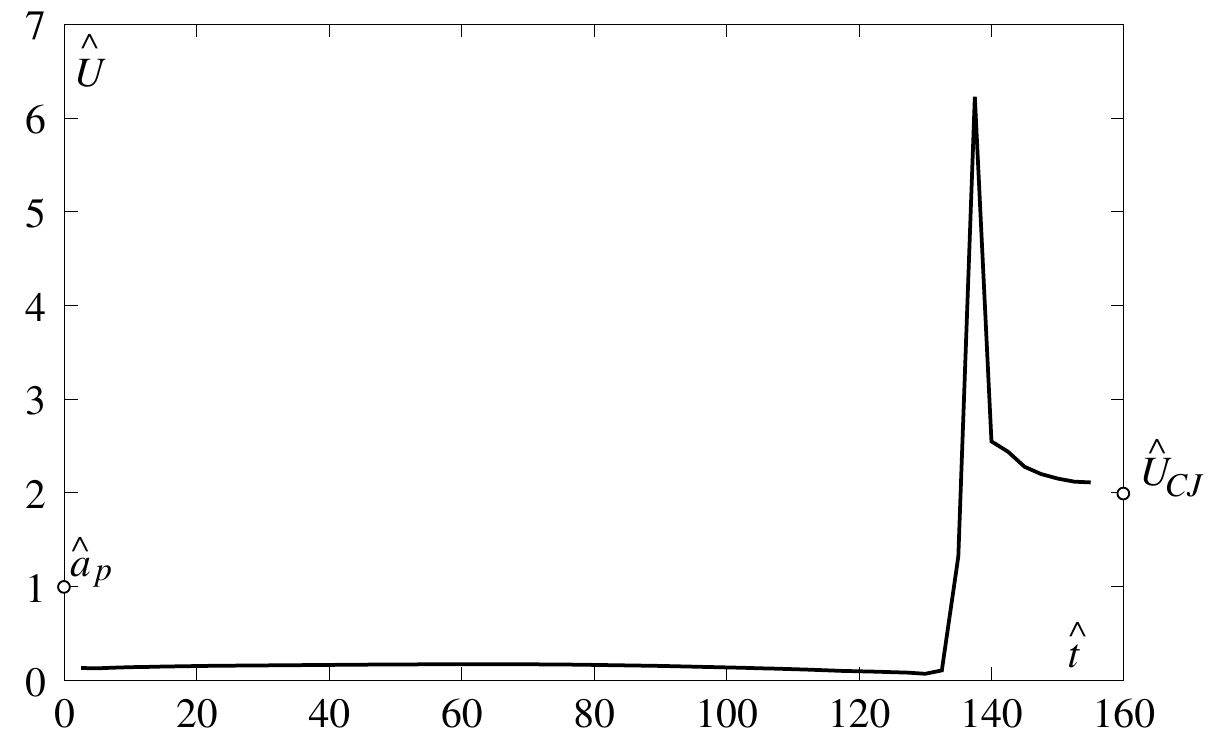}
\caption{
Reaction wave velocity $\hat{U}$ vs. time $\hat{t}$ along the system
centreline.
$\hat{U}_{CJ}$ corresponds to the Chapman -Jouguet detonation; $\hat{a}_p$ is
the velocity of sound in the burned mixture, $N_p = 3.5$, $\gamma$= 1.3.}
\end{figure}

\begin{figure}[!h]
\centering
\includegraphics[scale=0.75]{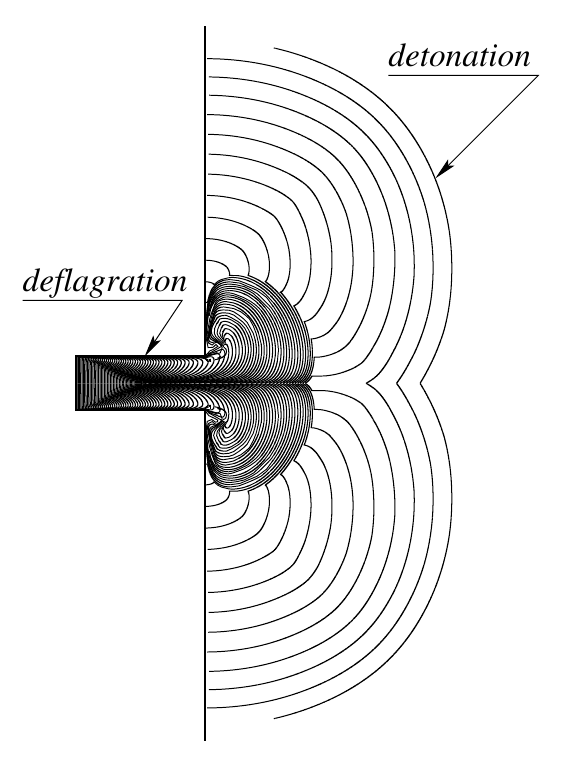}
\caption{
Detonation initiated at the exit from the tube, $N_p=4$, $\gamma=1.4$.} 
\end{figure}

\begin{figure}[!h]
\centering
\includegraphics[scale=0.55]{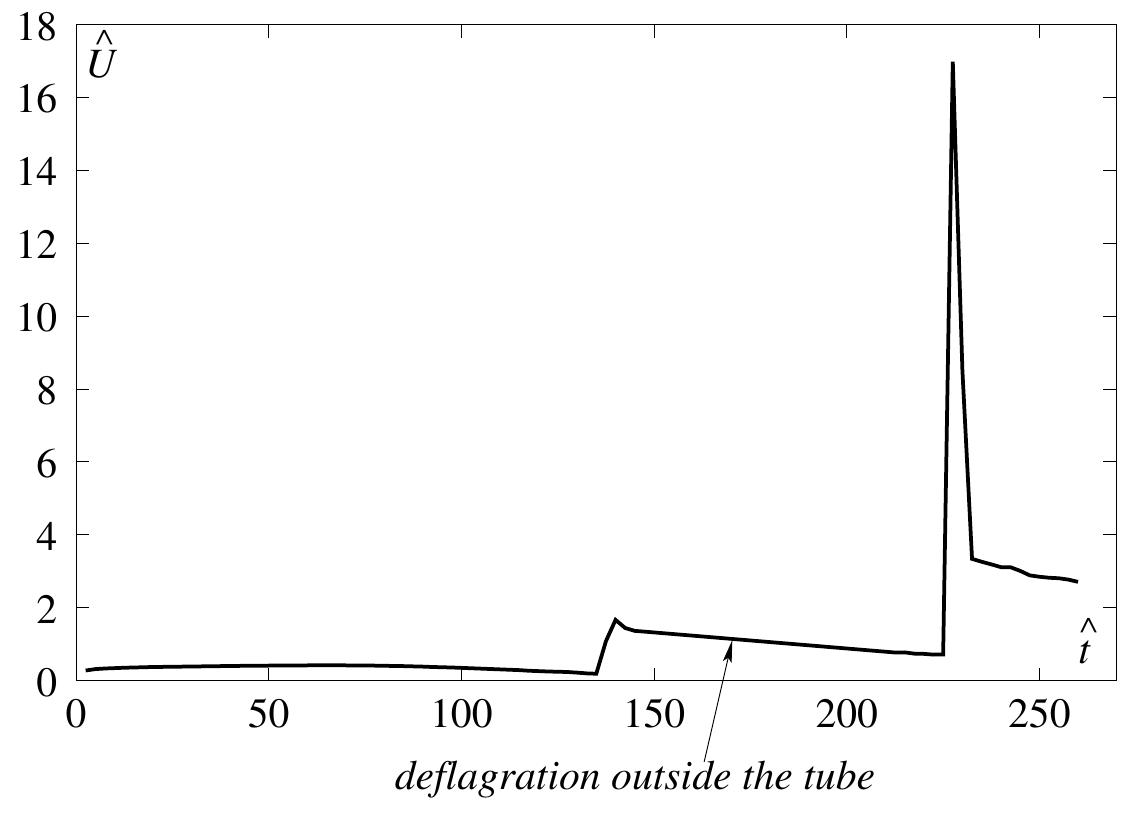}
\caption{
Reaction wave velocity $\hat{U}$ vs time $\hat{t}$ along the system
centreline; $N_p=4$, $\gamma=1.4$.  See also the caption of Fig. 3.}
\end{figure}

\begin{figure}[!h]
\centering
\includegraphics[scale=0.6]{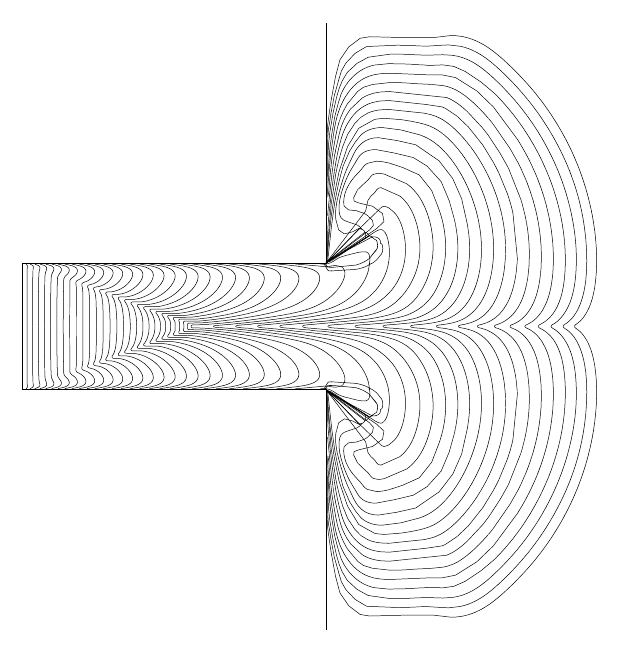}
\caption{
Deflagration exits the tube without transition to detonation;
$N_p=4$, $\gamma=1.3$.}
\end{figure}

\begin{figure}[!h]
\centering
\includegraphics[scale=0.55]{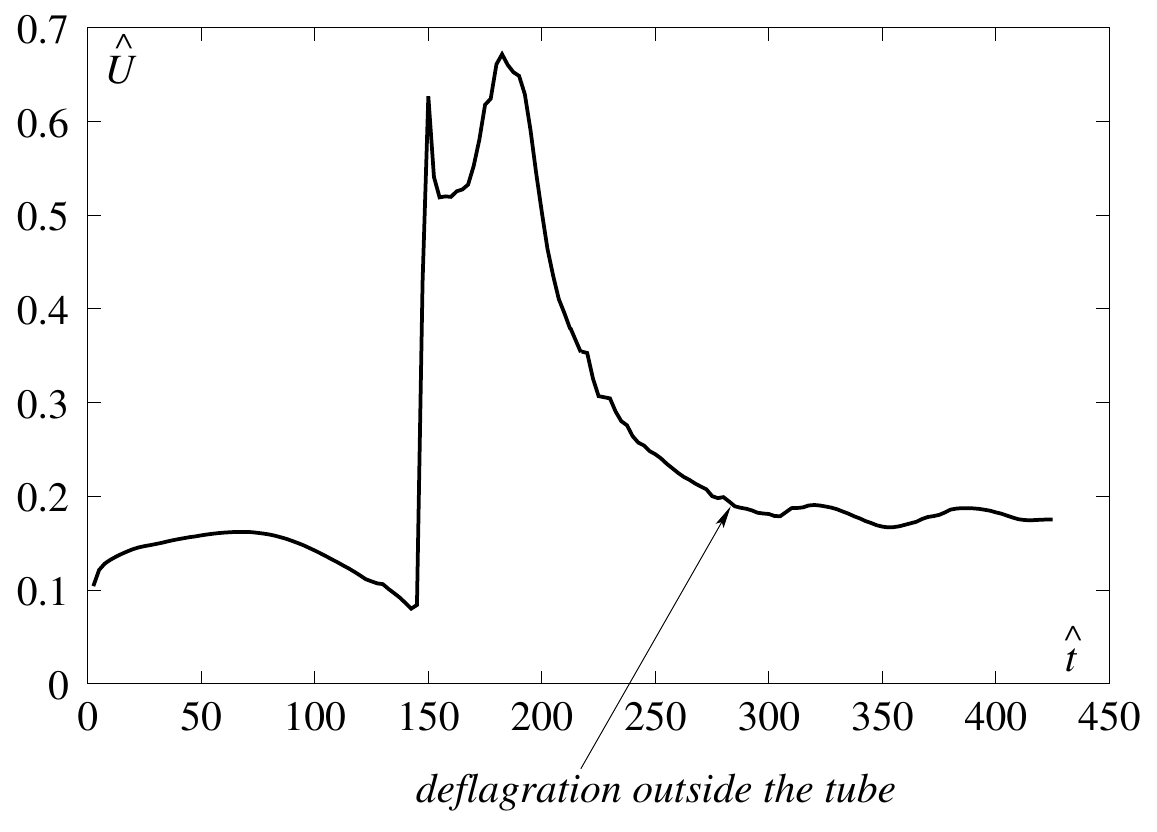}
\caption{
Reaction wave velocity $\hat{U}$ vs. time $\hat{t}$ along the system
centreline; $N_p=4$, $\gamma=1.3$.}
\end{figure}

The computational method used and its validation are described in [9].
Depending on parameters of the system one ends up with three scenarios:

1. Detonation initiated in the interior of the tube (Figs. 2, 3).

2. Detonation initiated at the exit from the tube (Figs. 4, 5)
- see also experimental Fig. 6 of Ref. [10].

3. Deflagration exits the tube without transition to detonation (Figs. 6,7).
Transition to detonation requires the tube to be long enough (Figs. 8,9).

\begin{figure}[h]
\centering
\includegraphics[scale=0.75]{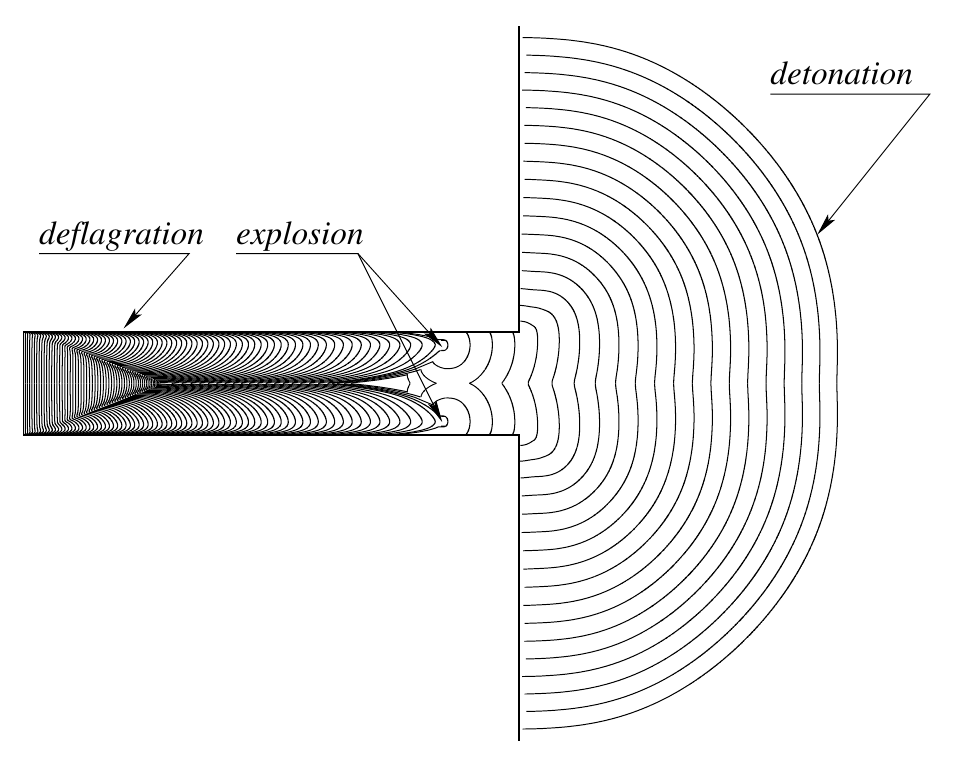}
\caption{\label{f:8}  Transition to detonation occers inside the tube;
$N_p=4$, $\gamma=1.3$, $[ab]=72$.}
\end{figure}

\section{Concluding remarks}

There are several promising directions for further studies of DDT in fuel-air clouds:

(a) to extend the above formulation from planar to circular geometry in order
to capture the incipient acceleration of the flame and reduction of the
predetonation time and distance, other parameters being fixed.

(b) to modify the geometry of the problem by removing vertical walls $gf$ and
$bc$, and thus allowing the emerging cylindrical/spherical detonation to touch
the tube's outer walls, $hg$, $ab$.

(c)  to reduce 2D, 3D models to two 1D models coupled at the exit from the
tube.  The first 1D model will be based on the Fanno hydraulic resistance
formulation [7], while the second will describe an expanding
cylindrical/spherical detonation in an unconfined space.

\clearpage

\begin{figure}[h]
\centering
\includegraphics[scale=0.55]{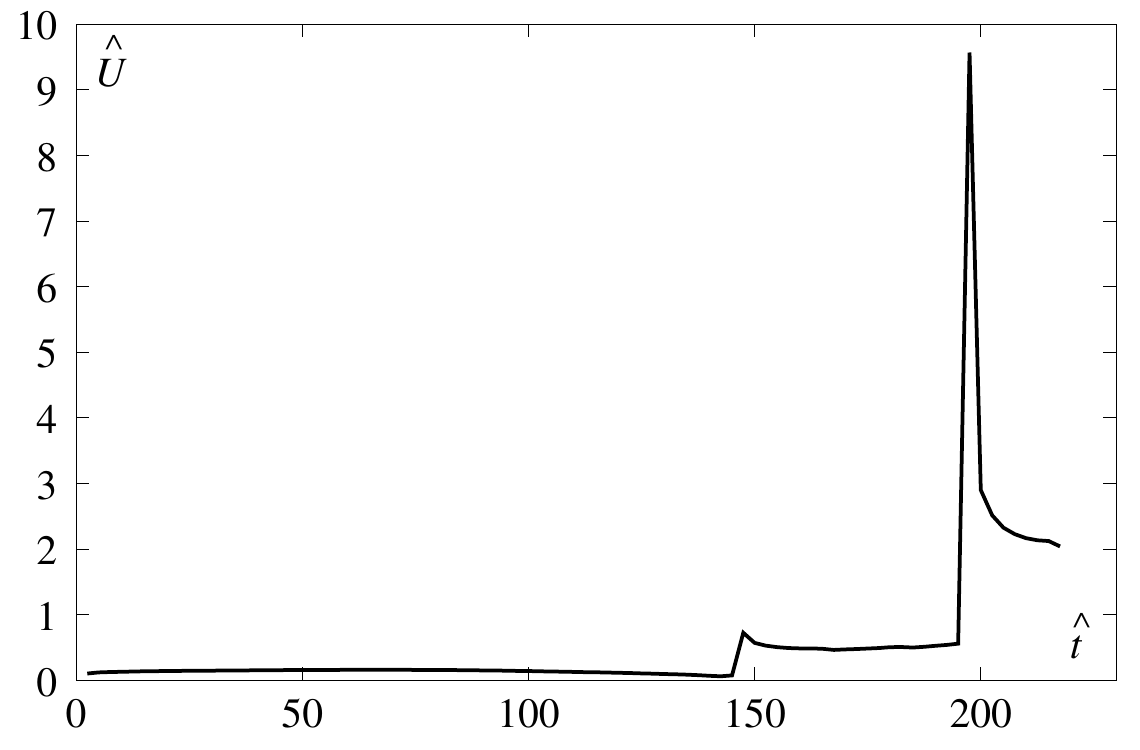}
\caption{ \label{f:9}
Reaction wave velocity $\hat{U}$ vs. time $\hat{t}$ along the system
centreline; $N_p=4$, $\gamma=1.3$, $[ab]=72$.}
\end{figure}

\section*{Acknowledgments}
These studies were supported by the US-Israel Binational Science Foundation
under Grant No. 2024033.

\end{document}